\documentclass[11pt]{article}
\usepackage{amsmath,amssymb,amsthm,mathrsfs}
\usepackage{array}
\usepackage{graphicx,psfrag,epsf}
\usepackage{multirow}
\usepackage{multicol}
\usepackage{multirow}
\usepackage{booktabs}
\usepackage{natbib}
\usepackage{appendix}
\usepackage{float}
\usepackage[colorlinks,citecolor=blue,linkcolor=blue,urlcolor=blue]{hyperref}

\newtheorem{theorem}{Theorem}

\newtheorem{corollary}{Corollary}

\begin{document}

\title{Cauchy combination test: a powerful test with analytic $p$-value calculation under arbitrary dependency structures}

\author{Yaowu Liu\thanks{Postdoctoral fellow, Department of Biostatistics, Harvard T.H. Chan School of Public Health, Boston, MA 02115} \;  and \; Jun Xie\thanks{Professor, Department of Statistics, Purdue University, West Lafayette, IN 47906. Research Sponsored by the National Institutes of Health Grant R21GM101504.}}

\date{}

\maketitle
\thispagestyle{empty}
\baselineskip=20pt

\begin{abstract}
Combining individual $p$-values to aggregate multiple small effects has a long-standing interest in statistics, dating back to the classic Fisher's combination test. In modern large-scale data analysis, correlation and sparsity are common features and efficient computation is a necessary requirement for dealing with massive data. To overcome these challenges, we propose a new test that takes advantage of the Cauchy distribution. Our test statistic has a very simple form and is defined as
a weighted sum of Cauchy transformation of individual $p$-values. We prove a non-asymptotic result that the tail of the null distribution of our proposed test statistic can be well approximated by a Cauchy distribution under arbitrary dependency structures. Based on this theoretical result, the $p$-value calculation of our proposed test is not only accurate, but also as simple as the classic $z$-test or $t$-test, making our test well suited for analyzing massive data. We further show that the power of the proposed test is asymptotically optimal in a strong sparsity setting. Extensive simulations demonstrate that the proposed test has both strong power against sparse alternatives and a good accuracy with respect to $p$-value calculations, especially for very small $p$-values. The proposed test has also been applied to a genome-wide association study of Crohn's disease and compared with several existing tests.
\end{abstract}

\noindent
 \textbf{Keywords:} Cauchy distribution; Correlation matrix; Non-asymptotic approximation; High dimensional data; Global hypothesis testing; Sparse alternative.

\pagestyle{plain}

\section{Introduction}

Methods for combining individual $p$-values or test statistics are of historically substantial interest in statistics.
A few well-known classical methods include
the Fisher's combination test~\citep{fisher1932statistical} and the sum-of-squares type tests. However, in modern high-throughput data analysis where there is only a small fraction of significant effects, these traditional tests are ineffective and can have substantial power loss~\citep{koziol1978combining,arias2011global}. For example, in genome-wide association studies (GWAS), massive amounts of genetic variants, e.g., single nucleotide polymorphism (SNP), are collected while only a small number of them are expected to be related to the phenotype of interest (e.g., a disease status).
Various methods have been developed to improve power for detecting sparse alternatives in this situation. The Tippett's minimum p-value test~\citep{tippett1931methods}, the higher criticism test~\citep{donoho2004higher}, and the Berk-Jones test~\citep{berk1979goodness} are particularly popular and have received substantial attention in the literature. As all three tests combine individual $p$-values to aggregate multiple effects, we will also refer to them as combination tests hereafter.

In practice, the individual test statistics or $p$-values are often correlated. For instance, SNPs could be highly correlated due to linkage disequilibrium. In order to control the Type I error and draw valid statistical inferences, the correlation structure should be taken into account in the $p$-value calculation. Here and throughout this paper, by $p$-value calculation, we always mean the $p$-value of a combination test, rather than the individual $p$-values that can be readily calculated. To the best of our knowledge, no analytic methods are available for the $p$-value calculation of the Tippett's minimum p-value, higher criticism, and Berk-Jones tests under dependence structures. While permutation or other approaches based on numerical simulations~\citep[e.g.,][]{liu2018accurate} can be used to incorporate the correlation information, they are computationally burdensome or even at times intractable in the analysis of massive data, especially in the following situations. First, when a combination test needs to be performed numerous times, such as in large-scale multiple testing, it is too time consuming to use the permutation approach. Here and throughout this paper, by large-scale multiple testing, we always mean a large number of combination tests instead of individual tests. Second, when the $p$-value of a combination test is extremely small, permutation would require very intensive computation as a vast number of simulations are needed to stabilize the calculation. This situation is particularly important in large-scale multiple testing, where practitioners care about the validity of extremely small $p$-values. Third, when the number of individual $p$-values in a combination test, denoted by $d$, is very large (e.g., $d=10^6$), the permutation approach is also impractical. Therefore, there is an increasing demand for developing tests whose $p$-values can be calculated analytically under dependence. Recently, \cite{barnett2017generalized} generalized the higher criticism test to incorporate the dependency structure and provided an analytic approximation method to compute the $p$-value of their new test. However, their analytic method is not accurate for extremely small $p$-values and still requires very intensive computation, even for a moderately large $d$ (e.g., $d>100$). In summary, due to a lack of computational efficiency and accuracy of $p$-values, none of the existing tests can handle large-scale data effectively.

Our main motivating examples of these challenging situations are also from GWAS, where the genetic data could contain hundreds of thousands of subjects and millions of SNPs and fast computation is a necessary requirement for the analysis of such big data.
One commonly used analysis approach in GWAS is to perform set-based analysis~\citep{wu2010powerful}, which divides the SNPs into sets/groups (e.g., genes) based on some biological information and tests the association between each SNP-set and the phenotype one at a time. The combination tests are useful for testing the significance of each SNP-set by aggregating the p-values of individual SNPs. In each SNP-set, the number of SNPs is not very large and often in dozens. However, there are tens of thousands of sets which need to be tested, requiring a fast calculation of the p-value for each SNP-set. In addition, a stringent significance threshold needs to be applied to account for multiple testing and the significant SNP-sets, which are the primary interest of practitioners, have very small p-values (e.g., $<10^{-6}$). Therefore, the p-value of a combination test needs to be accurate for exceedingly small p-values. Furthermore, one may be also interested in inferring whether there is any overall effect in the whole genome with millions of SNPs all together. This corresponds to the high-dimensional situation with a very large $d$.

In this article,
we propose a new combination test based on the Cauchy distribution and refer to it as the Cauchy combination test.  Similar to the Fisher's combination test, the new test statistic is defined as the weighted sum of transformed p-values~\citep{xie2011confidence,xie2013confidence}, except that the p-values are transformed to follow a standard Cauchy distribution.
We prove that the tail of the null distribution of this test statistic is approximately Cauchy under arbitrary correlation structures. According to this theoretical result, we then propose to calculate the $p$-value of the Cauchy combination test by the cumulative distribution function (c.d.f.) of a standard Cauchy distribution.
Similar to the classic $z$-test or $t$-test, our test has low computational requirements for $p$-value calculation and therefore is (potentially) able to be used routinely in large-scale data analysis with a vast number of combination tests.
We also establish similar theoretical result for the high-dimensional situation where the number of $p$-values $d$ diverges.
An extensive simulation study is carried out in Section~\ref{Sec:application}, which shows that under general correlation structures, the analytic $p$-value approximation by the Cauchy distribution is very accurate, especially for extremely small $p$-values. In fact, the smaller the $p$-value, the more accurate the approximation. In addition, parallel to the optimality theory for the minimum $p$-value test shown in \cite{arias2011global},
we prove that the power of our test is asymptotically optimal in a strong sparsity setting. In summary, the Cauchy combination test is well suited to deal with the challenges posed by sparsity, correlation, high-dimensionality and large scales, which, for example, are the situations we encountered in GWAS.

A related and more profound theory regarding the Cauchy distribution was also established in a recent work by~\cite{pillai2016unexpected}, which showed a remarkable result that the sum of some class of dependent Cauchy variables could be exactly Cauchy distributed. Our idea of using the Cauchy distribution was motivated from the strong need in GWAS for computationally scalable methods, and was originated from the observation that the sum of independent standard Cauchy variables follows the same distribution as the sum of perfectly dependent standard Cauchy variables.
We provide a detailed discussion on the connections and differences between ~\cite{pillai2016unexpected}'s and our work in Section~\ref{Subsec:whyCauchy}.

The rest of the paper is organized as follows. Section \ref{Sec:null} presents our main theorems about the null distribution of the Cauchy combination test statistic. In Section \ref{Sec:power}, we establish the asymptotical optimality of the power of the new test in the strong sparsity setting. In Section \ref{Sec:application}, we conduct extensive simulations to evaluate the accuracy of the $p$-value calculation of the proposed test and compare its power with a few existing tests. We use an analysis of GWAS data to demonstrate the effectiveness of our test. Some concluding remarks and a discussion of future research are given in Section \ref{Sec:discussion}. All the technical proofs and additional simulation results are relegated to the supplementary material.

\section{Null distribution} \label{Sec:null}

Let $p_i$ be the individual $p$-value, for $i=1,2,\cdots,d$. We define the Cauchy combination test statistic as
\begin{equation}\label{Sec1:TestStat}
T=\sum_{i=1}^d \omega_i \tan\{(0.5-p_i)\pi\},
\end{equation}
where the weights $\omega_i$'s are nonnegative and $\sum_{i=1}^d \omega_i=1$. Given that $p_i$ is uniformly distributed between 0 and 1 under the null, the component $\tan\{(0.5-p_i)\pi\}$ follows a standard Cauchy distribution.

When $p_i$'s are independent or perfectly dependent (i.e., all the $p_i$'s are equal), it is easy to see that the test statistic $T$ has a standard Cauchy distribution under the null. This phenomenon results from the closeness of Cauchy distribution under convolution and is unique to our Cauchy combination test statistic. In fact, for the minimum $p$-value, higher criticism, Berk-Jones, and many other test statistics, the null distribution in the independent case is completely different from that in the perfectly dependent case. This simple observation indicates that correlation can have a substantial impact on the null distribution of these existing tests and should not be ignored. While correlation also affects the null distribution of the Cauchy combination test statistic, we will show next that the impact on the tail is very limited.

\subsection{Non-asymptotic approximation for the null distribution}

To investigate the null distribution in the presence of correlation, we assume that the p-values are calculated from z-scores (i.e., test statistics that follow normal distributions).
Specifically, let $\mathbf{X}=(X_1,X_2,\cdots,X_d)^T$, where $X_i$ is a test statistic (or z-score) corresponding to the individual $p$-value $p_i$.

Denote $\text{E}(\mathbf{X})=\boldsymbol{\mu}$ and $\text{Cov}(\mathbf{X})=\boldsymbol{\Sigma}$. Because a test statistic must have a known null distribution to obtain its critical value or $p$-value, we can always rescale the individual test statistic $X_i$ to have variance 1. Thus, without loss of generality, we assume that $\boldsymbol{\Sigma}$ is a correlation matrix and then the individual $p$-value is given by $2\{1-\Phi(|X_i|)\}$. For testing the global null hypothesis that $H_0: \boldsymbol{\mu}=\mathbf{0}$,
we can rewrite the Cauchy combination test statistic \eqref{Sec1:TestStat} with respect to $\mathbf{X}$ as
\begin{equation*}
   T(\mathbf{X})=\sum_{i=1}^d \omega_{i}\tan[\{2\Phi(|X_i|)-3/2\}\pi].
\end{equation*}
Large values of $T(\mathbf{X})$ provide evidence against the global null hypothesis $H_0$.

We assume the following condition about the test statistics $X_i$'s.
\begin{enumerate}
  \item[(C.1)] (\textit{Bivariate normality}) For any $1\leq i< j\leq d$, $(X_i,X_j)^T$ follows a bivariate normal distribution.
\end{enumerate}
The bivariate normal condition (C.1) is mild and also assumed in~\cite{efron2007correlation}, where the author studied a similar topic, i.e., controlling the false discovery rate under dependence.

The following theorem provides a non-asymptotic approximation to the null distribution of $T(\mathbf{X})$ under any arbitrary correlation structure $\boldsymbol{\Sigma}$.
\begin{theorem}\label{MainTheorem}
  Suppose that the bivariate normality condition (C.1) holds and $E(\mathbf{X})=\mathbf{0}$. Then for any fixed $d$ and any correlation matrix $\boldsymbol{\Sigma}\geq 0$, we have
  \begin{equation*}
     \lim_{t\rightarrow +\infty} \frac{P\{T(\mathbf{X})> t\}}{P\{W_0>t\}}=1,
  \end{equation*}
where $W_0$ denotes a standard Cauchy random variable.
\end{theorem}

Theorem \ref{MainTheorem} indicates that the test statistic $T(\mathbf{X})$ still has approximately a Cauchy tail under dependency structures. Note that $T(\mathbf{X})$ is defined as a weighted sum of ``correlated" standard Cauchy variables. Roughly speaking, because of the heaviness of the Cauchy tail, the correlation structure $\boldsymbol{\Sigma}$ only has limited impact on the tail of $T(\mathbf{X})$.

As $p$-value corresponds to the tail probability of the null distribution, Theorem \ref{MainTheorem} suggests that we can use the standard Cauchy distribution to approximate the $p$-value of the test based on $T(\mathbf{X})$.
Let $t_\alpha$ be the upper $\alpha$-quantile of the standard Cauchy distribution, i.e., $P\{W_0>t_\alpha\}=\alpha$.
We define an $\alpha$-level test as
\begin{equation}\label{CCT}
  R_\alpha(\mathbf{X})=I\{T(\mathbf{X})> t_\alpha\}
\end{equation}
and refer to it as the Cauchy combination test, where $I(\cdot)$ is an indicator function.

Suppose that we observe $T(\mathbf{x})=t_0$. From the c.d.f. of standard Cauchy distribution, the $p$-value of the test can be simply approximated by
\begin{equation}\label{Pval:cal}
 \text{p-value}=1/2-(\arctan t_0)/\pi.
\end{equation}
Therefore, given the observed test statistic, the computation cost of calculating $p$-value is almost negligible, making the Cauchy combination test $R_\alpha(\mathbf{X})$ well suited for analyzing massive data. Furthermore, Theorem \ref{MainTheorem} guarantees that the approximation should be particularly accurate for very small $p$-values, which are of primary interest in large-scale multiple testing but very difficult to be calculated accurately.

Note that $P\{T(\mathbf{X})> t_\alpha\}$ represents the actual size, denoted by $s_\alpha$, of the test $R_\alpha(\mathbf{X})$. Theorem \ref{MainTheorem} can be equivalently stated as the ratio of the size to significance level converges to 1 as the significance level tends to 0, i.e.,
\begin{equation*}
\lim_{\alpha\rightarrow 0} \frac{s_\alpha}{\alpha}=1.
\end{equation*}
Simulation studies in Section~\ref{Subsec:simu:null} show that when the significance level $\alpha$ is moderately small, this ratio would already be close to 1 under a variety of correlation matrices.

Theorem~\ref{MainTheorem} can also be extended to the cases where the weights, $w_i$'s, are random and independent of the test statistics:
\begin{corollary} \label{Corollary:1}
  If the weights, $w_i$'s, are random variables and independent of $\mathbf{X}$, then Theorem~\ref{MainTheorem} still holds.
\end{corollary}

\subsection{Why Cauchy distribution?} \label{Subsec:whyCauchy}
In statistical literature, the Cauchy distribution mainly serves as a counter example, such as the nonexistence of the mean and an exception to the Law of Large Number. In this sense, quoting \cite{pillai2016unexpected}, ``some introductory courses have given the Cauchy distribution the nickname Evil". Probably for these reasons, the Cauchy distribution has seldom been used in statistical inference. Motivated by studying the large sample behavior of the Wald tests, \cite{pillai2016unexpected} recently revealed one of the angel aspects of Cauchy distribution and proved a surprising result that was originally conjectured by \cite{drton2016wald}. Specifically, let $\mathbf{Y}=(Y_1,\cdots, Y_d)^T$ and $\mathbf{Z}=(Z_1,\cdots, Z_d)^T$ be i.i.d. $N_d(\mathbf{0},\boldsymbol{\Sigma})$. Note that $Y_i/Z_i$ is Cauchy distributed. \cite{pillai2016unexpected} proved that for an arbitrary covariance matrix $\boldsymbol{\Sigma}$, $\sum_{i=1}^d \omega_i(Y_i/Z_i)$ still follows a standard Cauchy distribution, where $\sum_{i=1}^d \omega_i=1$ and $\omega_i\geq 0$ for any $i=1,\cdots d$.

Both their result and our Theorem~\ref{MainTheorem} indicates that the Cauchy distribution could be insensitive to certain types of dependency structures. Specifically, their result shows that the weighted sum of a class of dependent Cauchy variables can still be a Cauchy variable, while our Theorem~\ref{MainTheorem} considers another class of dependent Cauchy variables and indicates that the weighted sum of them still has a Cauchy tail. Therefore, we can use the Cauchy distribution to construct test statistics to deal with dependence structures, which are very challenging to be accounted for in general.

While sharing a similar interpretation with~\cite{pillai2016unexpected}, our result is substantially different and has unique contributions in terms of both methodology and theory. First and foremost, \cite{pillai2016unexpected}'s result is not a practical method and was motivated from a theoretical interest of studying the large-sample behaviour of the Wald test. After all, we rarely have a test statistic that is the ratio of two normal variables in practice. In contrast, the Cauchy combination test that we proposed maps the p-values to Cauchy variables and has a wide range of applications. Second, there is also fundamental distinctions between our and \cite{pillai2016unexpected}'s theories. Let $V_i=\tan[\{2\Phi(|X_i|)-3/2\}\pi]$ denote the Cauchy variable in our theory and $W_i=Y_i/Z_i$ denote the Cauchy variable in theirs, where $i=1,2,\cdots,d$. While $V_i$ and $W_i$ have the same marginal distribution, the joint distribution of $V_i$'s is completely different from that of $W_i$'s. Therefore, the test statistics $\sum\omega_iV_i$ and $\sum\omega_iW_i$ might follow very distinct distributions. In fact, our proof strategy for Theorem~\ref{MainTheorem} is also completely different from that in \cite{pillai2016unexpected}. In addition, because the bivariate normality condition assumed in our Theorem~\ref{MainTheorem} is much weaker than the joint normality assumption in~\cite{pillai2016unexpected}, our result allows for a broader class of dependent Cauchy variables than theirs. More discussions about the bivariate normality condition in high dimensions are provided in Section~\ref{Sec2:HighDimen}.

\subsection{Asymptotic approximation for the null distribution in high dimensions}\label{Sec2:HighDimen}

To establish the null distribution in the high dimensional situation where the number of p-values $d$ is very large and diverging, we further assume the following conditions on the correlation matrix $\boldsymbol{\Sigma}$.
\begin{enumerate}
  \item[(C.2)] $\lambda_{max}(\boldsymbol{\Sigma})\leq C_0$ for some constant $C_0>0$, where $\lambda_{max}(\boldsymbol{\Sigma})$ denotes the largest eigenvalue of $\boldsymbol{\Sigma}$.
  \item[(C.3)] $\max_{1\leq i< j\leq d} \sigma_{ij}^2\leq \sigma_{max}^2 < 1$ for some constant $0<\sigma_{max}^2 < 1$, where $\sigma_{ij}$ is the $(i,j)$-th element of $\boldsymbol{\Sigma}$.
\end{enumerate}
Conditions (C.2) and (C.3) on the correlation matrix are mild and common assumptions in the high dimensional setting~\citep[see, e.g.,][]{tony2014two}.

The following theorem shows that the Cauchy approximation for the null distribution is still valid when the dimension $d$ diverges.
\begin{theorem}\label{AsyptoticalNull}
  Suppose that conditions (C.1), (C.2) and (C.3) hold and $E(\mathbf{X})=\mathbf{0}$. If $d=o(t^c)$ for any constant $0<c<1/2$, we have
  \begin{equation*}
     \lim_{t\rightarrow +\infty} \frac{P\{T(\mathbf{X})> t\}}{P\{W_0>t\}}=1,
  \end{equation*}
where $W_0$ denotes a standard Cauchy random variable.
\end{theorem}

In addition to the theoretical justification provided in Theorem~\ref{AsyptoticalNull}, the Cauchy combination test also offers advantages that make it appealing in the high-dimensional situation from a practical point of view.
We illustrate the challenges and the advantages of the Cauchy combination test in the high-dimensional situation by the following example.
Let $\mathbf{Z}=(\mathbf{Z}_1,\mathbf{Z}_2,\cdots,\mathbf{Z}_d)$ be an $n\times d$ fixed design matrix and $\mathbf{Y}$ be a vector of $n$ i.i.d responses. Assume that $\mathbf{Y}$ and $\mathbf{Z}_i$'s are standardized to have mean 0 and variance 1. To test the marginal association between $\mathbf{Y}$ and $\mathbf{Z}_i$'s, we have the individual test statistics defined as $\mathbf{X}=\mathbf{Z}^T\mathbf{Y}/\sqrt{n}$.
Many classic tests use $X_i$ as the test statistic, such as the Cochran-Armitage trend test that is commonly used in GWAS for testing the association between a disease status and individual SNPs. When the $p$-values of all the SNPs in the genome are combined for a global significance, $d$ is in the hundreds of thousands or even millions.

First, the correlation matrix $\boldsymbol{\Sigma}$ is highly singular in the high dimensional situation, with a rank less than the sample size $n$ that could be much smaller than the dimension $d$. In the above example, $\boldsymbol{\Sigma}=\mathbf{Z}^T\mathbf{Z}/n$. For the minimum $p$-value, higher criticism, Berk-Jones, and many other tests, a highly singular correlation matrix would have a substantial impact on the null distribution and is very difficult to be accounted for. Note that perfect dependence is a special case of a highly singular correlation matrix, and that the Cauchy combination test statistic follows exactly a standard Cauchy distribution in this case. Thus, the Cauchy approximation should be particularly accurate in the high-dimensional situation. Our simulation study in Section~\ref{Subsec:simu:null} also confirms this expectation. Moreover, as the $p$-value of our test is calculated by a simple explicit formula~\eqref{Pval:cal} without requiring the information of $\boldsymbol{\Sigma}$ that could be very big in the high-dimensional situation, there is no computational issue for the Cauchy combination test even when $d$ is exceedingly large.

Second, the bivariate normality condition in Theorem~\ref{MainTheorem} and~\ref{AsyptoticalNull}~\citep[also in][]{efron2007correlation} is mild and appropriate in the high-dimensional setting. To see this, we compare it with the stronger condition of joint normality (i.e., $\mathbf{X}\sim N_d(\boldsymbol{\mu},\boldsymbol{\Sigma})$), which is commonly assumed in the literature when dependency structures are considered~\citep[see, e.g.,][]{hall2010innovated,fan2012estimating,fan2016estimation}. Consider the aforementioned example when the response is not normally distributed, such as a binary response. Both the bivariate and joint normality conditions are on the basis of multivariate central limit theorem (CLT). However, due to the convergence rate of CLT, $\mathbf{X}$ may not jointly converge to a multivariate normal distribution in the high-dimensional scenario where $d$ increases with $n$ at a certain rate. See \cite{chernozhukov2013gaussian} and \cite{chernozhukov2014central} for recent reviews on this topic. Therefore, it is not realistic to assume the joint normality of $\mathbf{X}$ when $d$ is comparable with or even much larger than $n$. The bivariate normality, however, is a much weaker assumption and still reasonable in the high-dimensional setting.

\subsection{Remarks}

{\bf Remark 1}. According to the test statistic \eqref{Sec1:TestStat} and $p$-value calculation \eqref{Pval:cal}, our test only requires the individual $p$-values (and the prespecified weights) as input. The bivariate normality condition and the correlation matrix $\boldsymbol{\Sigma}$ are only used to study the null distribution of the test statistic and are actually not needed for the application of our Cauchy combination test itself.

{\bf Remark 2}. The weights $\omega_i$'s add flexibility to incorporate possible domain knowledge to boost power. For example, in GWAS, the biological information of genetic variants (e.g., annotation) can be integrated to improve the analysis power~\citep{lee2014rare}. In comparison, to the best of our knowledge, the minimum $p$-value, higher criticism and Berk-Jones tests do not allow for incorporation of weights, as all of them have a maximum-type test statistic. In the absence of prior knowledge, the equal weights (i.e., $\omega_i=1/d$) can be employed.

{\bf Remark 3}. Because the null distribution of $T(\mathbf{X})$ is symmetric, i.e., $P\{T(\mathbf{X})> t\}=P\{T(\mathbf{X})< -t\}$ for any $t\in \mathbb{R}$, it is trivial that Theorem \ref{MainTheorem} also holds for the lower tail of the distribution of $T(\mathbf{X})$, i.e., $\lim_{t\rightarrow -\infty} P\{T(\mathbf{X})< t\}/P\{W_0<t\}=1.$

{\bf Remark 4}. As mentioned in Remark 1, the proposed test only takes the individual $p$-values as input.
Therefore, our method can also be useful in applications where the original data is difficult to access and only summary statistics, such as the individual test statistics or $p$-values, are available. For example, in GWAS, the original data can be difficult to obtain due to various reasons including consent and privacy issues. In fact, statistical analysis based on summary statistics has emerged with an increasing demand. Recently developed methods includes \cite{wen2010using,yang2012conditional,lee2014jepeg,finucane2015partitioning}.

{\bf Remark 5}. If the data are discrete and certain exact tests are used~\citep[see, e.g.,][]{liu2014exact}, the individual p-values may not exactly follow the uniform distribution $U[0,1]$ under the null. In many applications such as GWAS with a binary outcome~\citep{wu2010powerful,wu2011rare}, the p-values are often conservative, i.e., stochastically smaller than $U[0,1]$. Let $\mathbf{\tilde{X}}=(\tilde{X_1},\cdots,\tilde{X_d})$, where $\tilde{X_i}$ is a test statistic corresponding to the $p$-value $p_i$ and follows a normal distribution with mean 0 and variance less than 1. Then the p-value $p_i=2\{1-\Phi(|\tilde{X_i}|)\}$ is conservative. The following corollary shows that the Cauchy combination test can still protect the type I error and provide valid inference in the presence of conservative individual p-values.
\begin{corollary} \label{Corollary:2}
  Under the same assumptions of Theorem~\ref{MainTheorem} (or Theorem~\ref{AsyptoticalNull}) except that $\text{var}(\tilde{X}_i)\leq 1$ for any $i=1,2,\cdots,d$, we have
  \begin{equation*}
     \lim_{t\rightarrow +\infty} \frac{P\{T(\mathbf{\tilde{X}})> t\}}{P\{W_0>t\}}\leq1.
  \end{equation*}
\end{corollary}

\section{Power}\label{Sec:power}
 In this section, we study the asymptotic power of the proposed Cauchy combination test $R_\alpha(\mathbf{X})$ under sparse alternatives. Here asymptotics refers to $d$ tending to infinity.
 We follow the theoretical setup of \cite{donoho2004higher}. Assume that the individual test statistics $\mathbf{X}\sim N_d(\boldsymbol{\mu},\mathbf{\Sigma})$, where $\mathbf{\Sigma}$ is a banded correlation matrix, i.e., $\sigma_{ij}=0$ for any $|i-j|>d_0$ for some positive constant $d_0>1$. Let $\mu_i$ denote the coordinates of $\boldsymbol{\mu}$ for $i=1,2,\cdots,d$. We are interested in testing the global null hypothesis that $H_0: \boldsymbol{\mu}=\mathbf{0}$, against alternatives where only a small number of $\mu_i$'s are nonzero.
 Denote $S=\{1\leq i\leq d:\mu_i\neq 0\}$ as the set of signals or nonzero effects. Suppose that the number of signals $|S|=d^\gamma$, where $|S|$ is the cardinality of $S$ and the sparsity parameter $0<\gamma<1/2$. The signals are assumed to have the same magnitude, i.e., $|\mu_i|=\mu_0>0$ for all $i\in S$.

 \begin{theorem}\label{PowerTheorem}
    Suppose that $\min_{1\leq i\leq d}\omega_i\geq c_0/d$ for some constant $c_0>0$. Let $\mu_0=\sqrt{2r\log d}$, where $r>0$. For any $\alpha>0$, $r>(1-\sqrt{\gamma})^2$ and $0<\gamma<1/2$, we have
   \begin{equation*}
     \lim_{d\rightarrow +\infty} P\{ R_\alpha(\mathbf{X})=1 \}=1.
   \end{equation*}
 \end{theorem}

 Theorem \ref{PowerTheorem} states that the power of the Cauchy combination test converges to 1 for any significance level $\alpha>0$, or equivalently, that the sum of Type I and II errors can vanish asymptotically, under sparse alternatives. Furthermore, Theorem \ref{PowerTheorem} also indicates that the Cauchy combination test attains the optimal detection boundary defined in \cite{donoho2004higher} in the strong sparsity situation when $0<\gamma<1/4$.

 The power of our proposed test has some similarity to that of the minimum $p$-value test. In fact,~\cite{arias2011global} showed that the minimum $p$-value test also attains the optimal detection boundary when $0<\gamma< 1/4$. Intuitively, the minimum $p$-value test has good power against sparse alternatives since it uses the smallest individual $p$-value to represent the overall significance of a set of variables. In contrast, the Cauchy combination test statistic \eqref{Sec1:TestStat} transforms individual $p$-values to standard Cauchy variables. It can be easily seen that small $p$-values correspond to very large values of a Cauchy variable and the sum in \eqref{Sec1:TestStat} is essentially dominated by a few of the smallest $p$-values (see a toy example in Table 1 in the supplementary material). Roughly speaking, the Cauchy combination test uses the few smallest $p$-values to represent the overall significance. Therefore, it is expected to also have strong power against sparse alternatives.

 The theoretical analysis here only states the asymptotic power under banded correlation matrices. To examine the finite-sample power performance under general correlation structures, extensive simulation studies are carried out in Section~\ref{Subsec:simu:power} and show that the Cauchy combination test has very robust power across a range of correlation structures and sparsity levels compared with the existing tests. We also provide some discussions about the finite-sample power of the Cauchy combination test in Section 3 in the supplementary material.

\section{Applications}\label{Sec:application}
We evaluate the $p$-value calculation accuracy of the Cauchy combination test under a variety of hypothetical and real-data-based correlation matrices.
We also compare the power of our proposed test with three existing tests that have strong power against sparse alternatives, i.e., the minimum $p$-value~\citep{tippett1931methods}, higher criticism~\citep{donoho2004higher} and Berk-Jones~\citep{berk1979goodness} tests. Throughout this section, the weights, $\omega_i$'s, in the Cauchy combination test statistic are chosen to be $1/d$ for all $i=1,2,\cdots,d$.

For both real-data analysis and parts of the simulations, we use
the data of a Crohn's disease genome-wide association study~\citep{duerr2006genome}, which aims at identifying SNPs or genes that are associated with the inflammatory bowel disease.
This data contains 1028 independent subjects from the Non-Jewish population.
After similar data quality control as in \cite{duerr2006genome}, the data set used in our analysis consists of 293,426 SNPs and 997 subjects, with 498 cases and 499 controls. SNPs are grouped into 15,279 genes on chromosomes 1--22 according to the Genome Build UCSC hg17 assembly. The gene size (number of SNPs) ranges from 1 to 705 and is highly skewed to the right.

\subsection{Accuracy of $p$-value calculation} \label{Subsec:simu:null}
We use simulations to examine the accuracy of $p$-value calculation based on the Cauchy approximation under various correlation structures and different dimensions. The vector of individual test statistics $\mathbf{X}$ is generated from $N_d(0,\boldsymbol{\Sigma})$ under the null hypothesis. We consider six values of the dimension $d$, i.e., $d=5,20,50,100,300,500$, for each of the following correlation matrix $\boldsymbol{\Sigma}=(\sigma_{ij})$.

\begin{itemize}
  \item Model 1 (AR(1) correlation): For each $d$, $\sigma_{ij}=\rho^{|i-j|}$ for $1\leq i,j\leq d$, where $\rho=0.2,0.4,0.6,0.8,0.99$. There are 30 conditions in total, corresponding to six dimension values and five correlation matrices.
  \item Model 2 (Polynomial decay): For each $d$, $\sigma_{ii}=1$ and $\sigma_{ij}=\frac{1}{0.7+|i-j|^\rho}$ for $1\leq i\neq j\leq d$, where $\rho=0.5,1.0,1.5,2.0,2.5$. There are 30 conditions in total.
  \item Model 3 (Singular matrices): For each $d$, let $A=(a_{ij})$ be a $(d/5)\times d$ matrix, where $a_{ij}=\rho^{|i-j|}$ and $\rho=0.2,0.4,0.6,0.8,0.99$. Further let $D=(d_{ij})$ be a diagonal matrix with diagonal elements $d_{ii}=(\tilde{a}_{ii})^{-1/2}$, where $\tilde{a}_{ii}$ is the $i$-th diagonal of $A^TA$. Then we take $\boldsymbol{\Sigma}=D^TA^TAD$. There are 30 conditions in total.
  \item Model 4 (Real genotypes): For each $d$, we randomly select 10 genes from the Crohn's disease data that have or approximately have $d$ SNPs. Then we take $\boldsymbol{\Sigma}$ to be the sample correlation matrix of SNPs in a gene. There are 60 conditions in total.
\end{itemize}

Model 1 and 2 are commonly used in simulations. The singular matrices constructed in Model 3 aim to mimic the high-dimensional situation and contain many large and moderate correlation coefficients. We also consider realistic correlation structures in genetic data through Model 4. Since SNPs could be highly correlated due to linkage disequilibrium, the correlation matrices in Model 4 often contain very strong correlations (e.g., $0.99$).

Recall that our Theorem \ref{MainTheorem} indicates that $\lim_{\alpha\rightarrow 0}s_\alpha/\alpha=1$, where $s_\alpha$ denotes the size of the Cauchy combination test. For each correlation matrix $\boldsymbol{\Sigma}$ specified above, we generate $10^8$ Monte Carlo samples to evaluate the empirical size at significance levels $\alpha=10^{-1},10^{-2},10^{-3},10^{-4},10^{-5}$, and use the ratio of empirical size to significance level to reflect the accuracy of $p$-value calculation.

\begin{figure}[!h]
  \centering
  \includegraphics[scale=0.8]{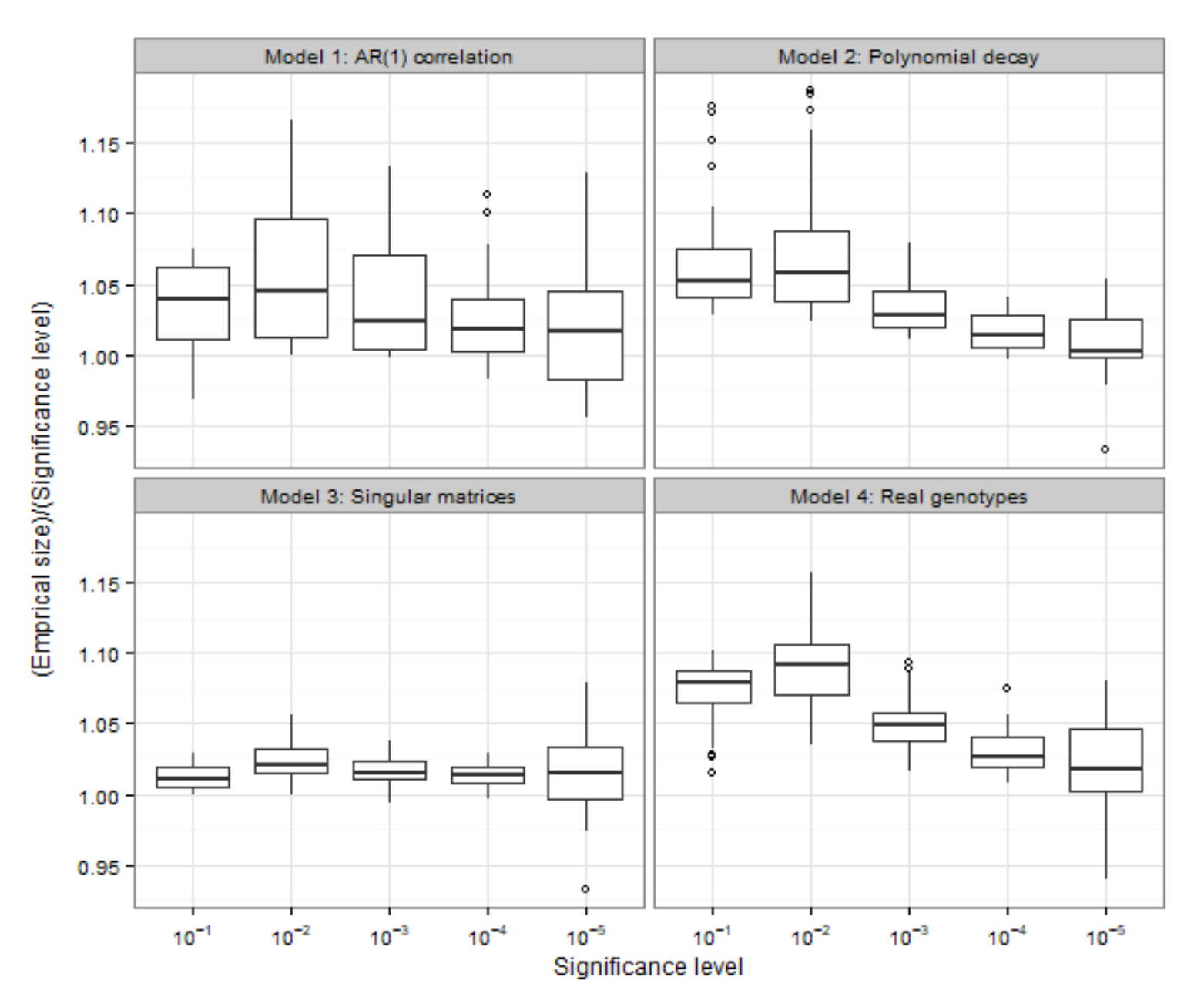}
  \caption{The ratio of empirical size to significance level for Model 1--4, summarized by boxplots. The $x$-axis is the significance level at $\alpha=10^{-1},10^{-2},10^{-3},10^{-4},10^{-5}$.}
  \label{Boxplots}
\end{figure}

The results are summarized by boxplots and shown in Figure \ref{Boxplots}. It can be seen that the Type I error of the Cauchy combination test is well controlled in general. As the significance level decreases, the Cauchy approximation becomes more accurate. For very small significance levels such as $\alpha=10^{-5}$, the Monte Carlo errors are not negligible and are in fact the main cause of the variations in the boxplots. Given the total number of our simulation conditions (i.e., 150), the ratio of empirical size to significance level for $\alpha=10^{-5}$ is not significantly different from 1, which indicates very good accuracy for extremely small $p$-values.
Furthermore, under real correlation structures in Model 4 that contain very strong correlations, the Type I error is still well controlled. This is expected because the perfect dependency is also an extreme case of strong correlation.
Moreover, it can be seen from the result of Model 3 that the accuracy is particulary good under singular correlation matrices, which agrees with our discussion in Section~\ref{Sec2:HighDimen}.

\begin{figure}[!h]
  \centering
  \includegraphics[scale=0.6]{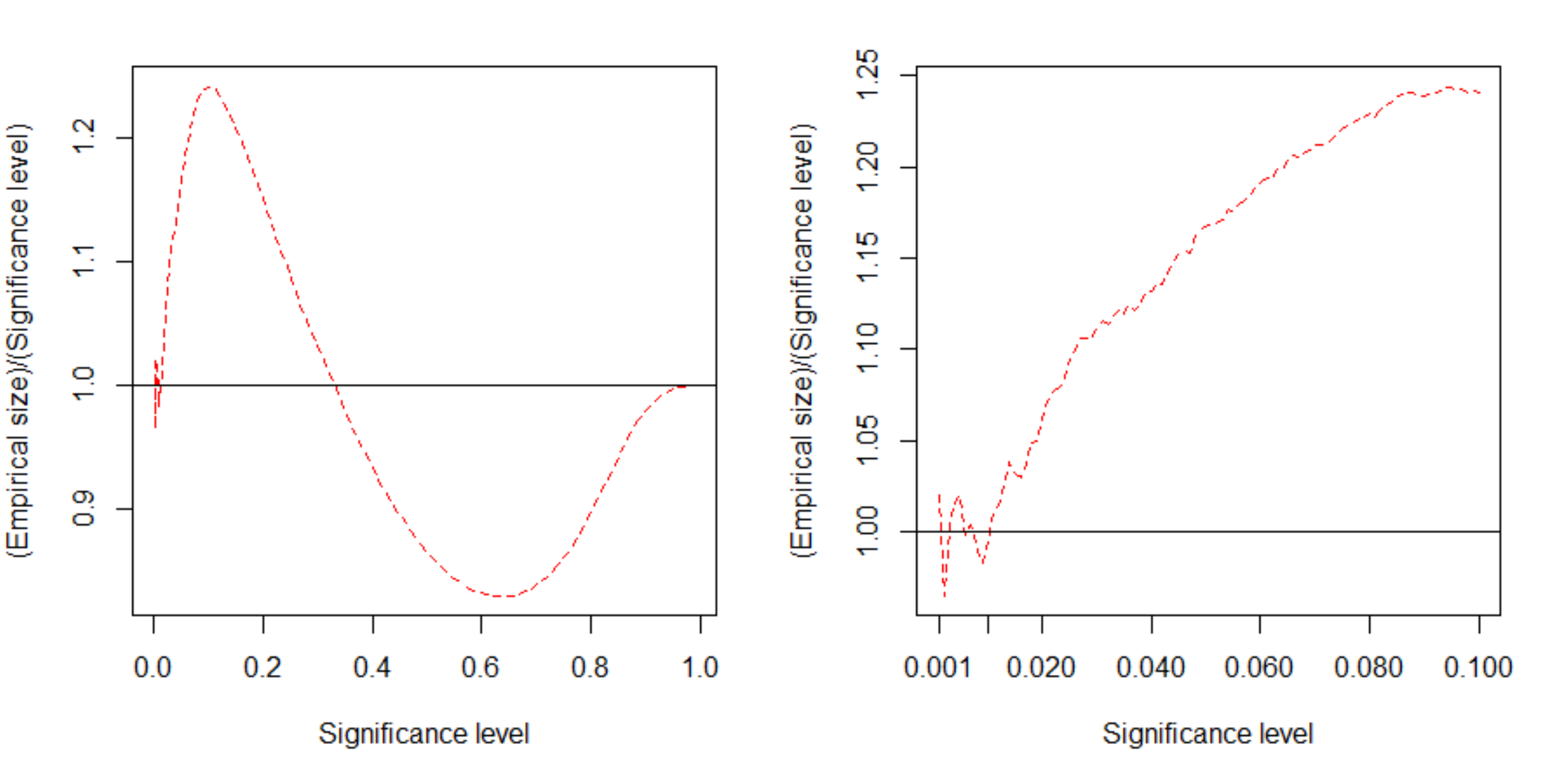}
  \caption{The ratio of empirical size to significance level (dashed lines) for Model 5. The straight line in each plot is the reference line. The plot on the right is a zoom-in image of the plot on the right. Note that the non-smoothness and fluctuation of the dashed curve in the right plot is due to the Monte Carlo errors.}
  \label{Null:whole:genome}
\end{figure}

To examine the accuracy of the $p$-value calculation in a more challenging high-dimensional scenario with an exceedingly large $d$, we consider the following correlation matrix $\boldsymbol{\Sigma}$ based on real genomic data.
\begin{itemize}
  \item Model 5 (High-dimensional singular matrix): We take $\boldsymbol{\Sigma}$ to be the sample correlation matrix of all the SNPs in the Crohn's disease data. Specifically, $\boldsymbol{\Sigma}$ is a highly singular matrix with dimension $d=293,426$ and rank equal to the sample size $n=997$.
\end{itemize}
Because the computation for a large $\boldsymbol{\Sigma}$ is very intensive, we use $10^6$ Monte Carlo samples to calculate the empirical sizes. Figure \ref{Null:whole:genome} shows the simulation result and demonstrates that the $p$-value calculation is still accurate under high-dimensional singular correlation matrix.

We also investigate the accuracy of $p$-value calculation when the normality assumption is violated. The simulation setting is exactly the same as that of Figure~\ref{Boxplots}, except that $\mathbf{X}$ is generated from a multivariate t distribution with 4 degrees of freedom, i.e, $\mathbf{X}\sim t_4(0,\boldsymbol{\Sigma})$. The result is presented in Figure 1 in the supplementary material and shows a similar phenomena as the Gaussian case.

\subsection{Power comparison}\label{Subsec:simu:power}

We compare the power of the Cauchy combination, minimum $p$-value, higher criticism, and Berk-Jones tests, which are denoted by \textit{CCT}, \textit{MinP}, \textit{HC} and \textit{BJ}, respectively.
More specifically, we investigate how sparsity and correlation could influence the power of different tests. Under the alternative, the vector of individual test statistic $\mathbf{X}$ is generated from $N_d(\boldsymbol{\mu},\boldsymbol{\Sigma})$, where $\boldsymbol{\mu}=(\mu_i)$ and $\boldsymbol{\Sigma}=(\sigma_{ij})$. Three values of the dimension $d$ are examined: $d=20,40,60$. The percentage of signals (i.e., non-zero $\mu_i$'s) is set to be $5\%$, $10\%$ and $20\%$ for each $d$. All the signals have the same strength $\mu_0$, which is chosen to be $\sqrt{3\log(d)}/s^{1/3}$ to make the power in different settings comparable, where $s$ denotes the number of signals.
The correlation matrix $\boldsymbol{\Sigma}$ is set to have an exchangeable structure, with $\sigma_{ij}=\rho$ for all $1 \leq i\neq j \leq d$, and a variety of correlation levels are considered with $\rho$ chosen to be the nonnegative multiples of 0.05 between 0 and 0.4.

For each $\boldsymbol{\Sigma}$, we first draw $10^5$ Monte Carlo samples to calculate the critical values of \textit{CCT}, \textit{MinP}, \textit{HC} and \textit{BJ} at the significance level $0.05$. Here we also use simulation-based critical values for \textit{CCT} in order to make a fair comparison. Then in each sparsity and correlation setting, $10^4$ simulations are performed to calculate the power of the four tests.

\begin{figure}[!h]
  \centering
  \includegraphics[scale=0.80]{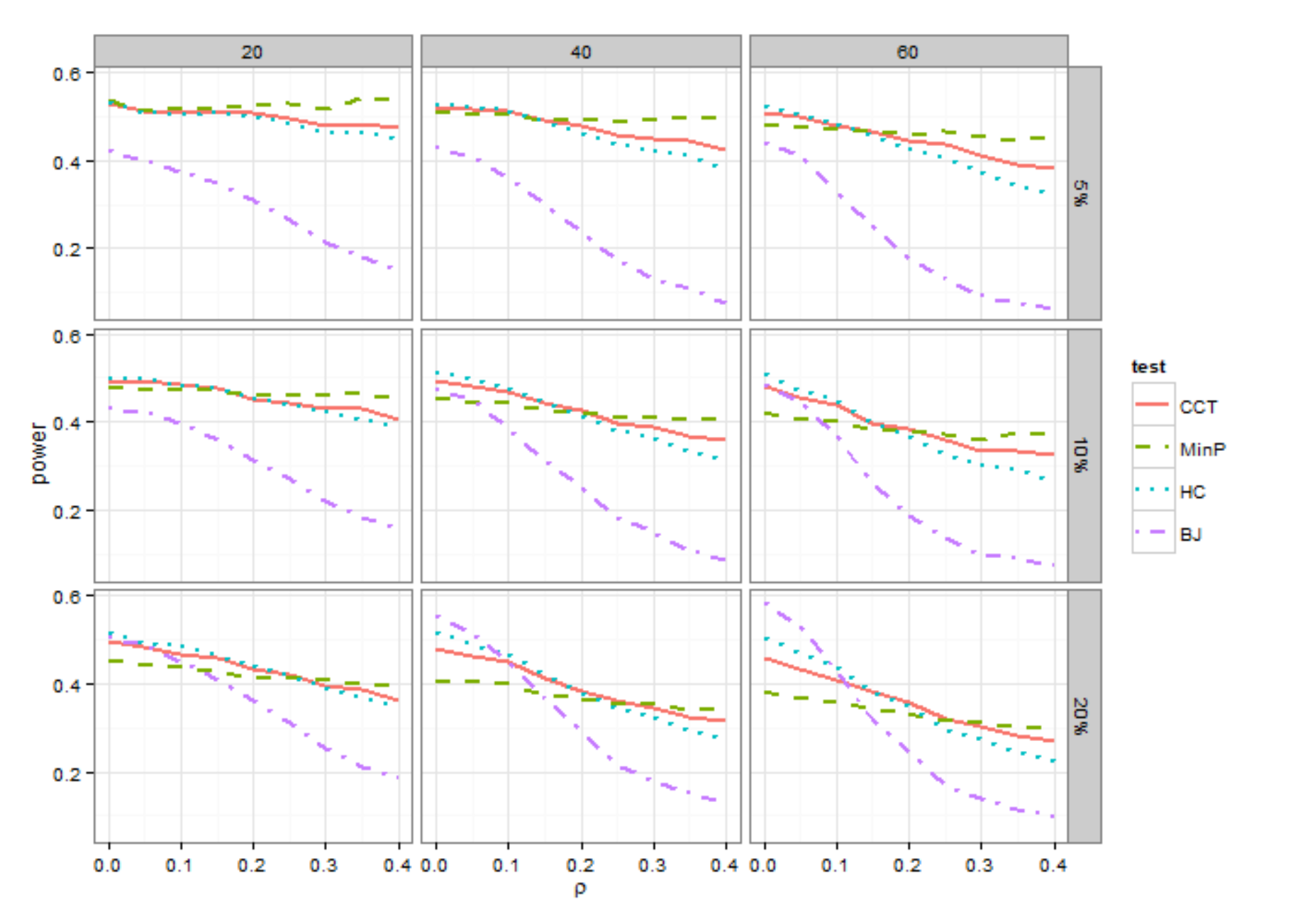}
  \caption{Power comparison of \textit{CCT}, \textit{MinP}, \textit{HC} and \textit{BJ}. The $x$-axis is the correlation strength $\rho$. The columns from left to right correspond to the dimension $d=20,40,60$. The rows from top to bottom correspond to the signal percentage $5\%$, $10\%$ and $20\%$. The signal strength is chosen to make the power in every setting comparable.}
  \label{Power:exchangeable}
\end{figure}

The result is displayed in Figure~\ref{Power:exchangeable}. It demonstrates that \textit{CCT} has very robust power across different sparsity and correlation levels compared with the other three tests. \textit{MinP} is not sensitive to the magnitude of correlation. But when signals are not very sparse and weakly dependent, \textit{MinP} has a considerable power loss and \textit{BJ} is most advantageous in this situation. Both \textit{HC} and \textit{BJ} lose power substantially as correlation increases, even in the case of moderately sparse signals. One possible explanation for this is that both tests compare the ordered individual $p$-value $p_{(i)}$ with the reference value $i/d$, which is not a correct reference in the presence of correlation. In contrast, \textit{CCT} has very robust power in every setting. It outperforms \textit{MinP} when signals are not very sparse, and has higher power than \textit{HC} and \textit{BJ} in the case of moderate or strong correlations. In the absence of prior knowledge of sparsity and correlation, such as scanning genes in GWAS, \textit{CCT} would be a robust choice and less likely to miss important signals. More importantly, the $p$-value of \textit{CCT} can be computed accurately and analytically under general correlation structures, while the other three tests would require intensive computation and are not suitable to analyze large-scale data. We also present the result based on analytic critical values of \textit{CCT} in Figure 2 in the supplementary material, which demonstrates a similar phenomena.

\subsection{Real genetic data analysis}
We apply our Cauchy combination test to the Crohn's disease genome-wide association study and compare it with the other three tests (i.e., \textit{MinP}, \textit{HC} and \textit{BJ}) in terms of power and computation time. All the analyses are carried out
on a computer node with 2.5 GHz quad-core Intel Xeon E3-1284 CPUs and 32 GB memory.

Firstly, we perform the single-SNP analysis. The individual $p$-values of the 293,426 SNPs in the study are obtained based on the Cochran--Armitage trend test for the association between the disease status and individual SNPs.
Two SNPs are found to be significant at a level of $0.05$ after the Bonferroni adjustment, with $p$-values of $2.8\times 10^{-8}$ and $7.5\times 10^{-8}$. The analysis is performed using the standard software \textit{Plink}~\citep{purcell2007plink} for genome-wide association studies and takes about 3 minutes. Then we exclude these two SNPs and apply our proposed test to combine the individual $p$-values of the remaining 293,424 SNPs. The Cauchy combination test gives a $p$-value of $0.030<0.05$, which suggests that there still exists genetic information in the remaining SNPs. The computation of this step only takes about 1 second. Note that the Cochran--Armitage trend test statistics have a correlation matrix $\boldsymbol{\Sigma}$ equal to the sample correlation matrix of SNPs. The result for Model 5 (Figure \ref{Null:whole:genome}) indicates that the $p$-value maintains satisfactory accuracy. In comparison, for the other three tests, permutation is needed to incorporate the correlation structure $\boldsymbol{\Sigma}$ to obtain accurate $p$-values. Because of the high dimensionality in this situation, it is computationally very intensive to use permutation and we do not provide the results for the other three tests here.

We next perform a gene-based analysis, i.e., the $p$-values of individual SNPs in a gene are combined to test for an overall significance of the gene. We apply the Cauchy combination test and the other three tests (i.e., \textit{MinP}, \textit{HC} and \textit{BJ}) to screen the 15,279 genes in the Crohn's disease data. The most significant genes based on the four tests are shown in Table \ref{tab:pval}. The $p$-values of our test are obtained according to \eqref{Pval:cal} and permutation is employed for computing the $p$-values of the other three tests. In particular, we use $10^8$ permutations for the genes listed in Table \ref{tab:pval} and $10^6$ permutations for the rest of genes. All four tests identify genes \textit{IL23R} and \textit{NOD2} as significant at a level of $0.05$ after the Bonferroni adjustment. These two genes are also found to contain genetic variants associated with the Crohn's disease in the literature~\citep{franke2010genome}. The proposed Cauchy combination test only takes about 10 seconds to complete the analysis, i.e., computing the $p$-values of all the genes, compared with nearly 12 days for the other three tests based on permutation. In addition, our simulation result of Model 4 in Section \ref{Subsec:simu:null} indicates that the $p$-values of the genes in Table \ref{tab:pval} based on the Cauchy combination test should be very accurate, since these genes have very small $p$-values.

\begin{table}[!h]
  \caption{\label{tab:pval} $P$-values of the most significant genes in analysis of the Crohn's disease data
  using the four tests. The list is sorted in increasing order based on the smallest of the
   $p$-values of the Cauchy combination test (\textit{CCT}).}
  \centering
  \fbox{
  \begin{tabular}{cccccc}
    Gene & $d$ & MinP  & HC    & BJ    & CCT \\
    \hline
    NOD2  & 8     & $2.00\cdot 10^{-7}$ & $1.80\cdot 10^{-7}$ & $1.20\cdot 10^{-7}$ & $4.35\cdot 10^{-7}$ \\
    IL23R & 22    & $1.20\cdot 10^{-7}$ & $1.60\cdot 10^{-7}$ & $3.10\cdot 10^{-7}$ & $5.84\cdot 10^{-7}$ \\
    OR2AT4 & 5     & $1.10\cdot 10^{-4}$ & $8.26\cdot 10^{-5}$ & $9.31\cdot 10^{-5}$ & $6.05\cdot 10^{-5}$ \\
    RASD2 & 22    & $6.86\cdot 10^{-5}$ & $8.95\cdot 10^{-5}$ & $9.21\cdot 10^{-4}$ & $8.45\cdot 10^{-5}$ \\
    SLCO2B1 & 16    & $1.53\cdot 10^{-4}$ & $6.96\cdot 10^{-5}$ & $1.42\cdot 10^{-4}$ & $1.22\cdot 10^{-4}$ \\
    VSX2  & 5     & $1.80\cdot 10^{-4} $& $1.82\cdot 10^{-4}$ & $1.95\cdot 10^{-3}$ & $1.98\cdot 10^{-4} $\\
    TIFA  & 4     & $1.66\cdot 10^{-4}$ & $2.38\cdot 10^{-4}$ & $7.27\cdot 10^{-3}$ & $2.00\cdot 10^{-4}$ \\
    SLC44A4 & 6     & $3.78\cdot 10^{-4} $& $1.70\cdot 10^{-4}$ & $1.00\cdot 10^{-3}$ & $2.40\cdot 10^{-4}$ \\
    GIMAP7 & 4     & $5.65\cdot 10^{-4}$ & $7.70\cdot 10^{-4}$ & $5.23\cdot 10^{-5}$ & $5.49\cdot 10^{-4}$ \\
    EVI5L & 7     & $2.35\cdot 10^{-2} $& $3.18\cdot 10^{-3}$ & $7.01\cdot 10^{-5}$ & $8.96\cdot 10^{-3}$ \\
  \end{tabular}}
\end{table}

In summary, for either single-SNP or gene-based analysis, our method can be done within just a few seconds and provide reasonably accurate $p$-values, while the other three existing tests are computationally burdensome for the analysis of large genomic data.

\section{Discussion}\label{Sec:discussion}

In this paper, we use the Cauchy distribution to construct a novel test that not only is powerful against sparse alternatives but also has accurate and efficient $p$-value calculations under arbitrary dependency structures. Our contributions are threefold. Firstly, the proposed Cauchy combination test fills the gap of testing against sparse alternatives.
In the case of dense signals, with a variety of analytic $p$-value calculation methods, the classical sum-of-squares tests have been widely used in practice. However, in the case of sparse signals, none of the existing tests have efficient $p$-value calculations, which are of great importance for analyzing massive or big data.
  Second, the analytic method for computing the $p$-value of our proposed test maintains several notable properties, making the test particularly useful in modern large-scale and high-dimensional data analysis. Finally, besides the methodological contribution, our Theorem~\ref{MainTheorem} also has interest on its own. It can be viewed as an extension of the closeness of Cauchy distribution under convolution from the independent case to a special dependent case. It is also established under very weak assumptions, essentially requiring only the bivariate normality of the individual test statistics.

The Cauchy combination test statistic $T$ in~\eqref{Sec1:TestStat} can be viewed as a special case of the general combination scheme based on the sum of transformed p-values~\citep{xie2011confidence,xie2013confidence}, i.e., $\sum_{i=1}^d h(p_i)$, where $h(\cdot)$ can be any monotonically increasing function. Besides the advantages resulting from the special Cauchy transformation, this general combination scheme has many other advantages, for example, being able to make an exact inference in discrete data analysis and enhance finite sample efficiency~\citep{liu2014exact}. It is of great interest to explore other transformations and use this general combination scheme to develop tests that have other remarkable features.

While the bivariate normality assumption in Theorem~\ref{MainTheorem} is appropriate in many cases, there are some applications where the individual $p$-values are calculated from test statistics that are not normally distributed and one can also apply the proposed test to combine the $p$-values.  We observe through simulations that the Cauchy approximation is still quite accurate in these situations where the normality assumption is not satisfied, for example, the simulation under multivariate t distribution in Figure 1 in the supplementary material.
Hence, it is interesting to generalize Theorem~\ref{MainTheorem} to non-Gaussian individual test statistics.

The accuracy of the Cauchy approximation would certainly depend on the significance level and the correlation structure, which we have investigated empirically. Another interesting research question is to derive the convergence rate or a non-asymptotic bound for Theorem~\ref{MainTheorem}.

\section*{Acknowledgments}
The authors thank the associate editor and the two anonymous referees for their
comments and suggestions that have helped greatly improve the
paper. The first author would like to thank  Xihong Lin for multiple inspiring discussions.

\section*{Supplementary material}
Supplementary material includes the proofs of Theorem~\ref{MainTheorem}--\ref{PowerTheorem}, Corollary~\ref{Corollary:1}--\ref{Corollary:2} and technical lemmas, additional simulation results, as well as some further discussions about the finite-sample power of the proposed test.

\bibliographystyle{chicago}
\bibliography{Cauchy_Test}

\end{document}